# Spin-flop transition in atomically thin MnPS$_3$ crystals

Gen Long[1,2,5], Hugo Henck[1,2,5], Marco Gibertini[1,3], Dumitru Dumcenco[1], Zhe Wang[1,2], Takashi Taniguchi[4], Kenji Watanabe[4], Enrico Giannini[1], Alberto F. Morpurgo[1,2*]

1. Department of Quantum Matter Physics, University of Geneva, 24 Quai Ernest Ansermet, CH-1211 Geneva, Switzerland
2. Group of Applied Physics, University of Geneva, 24 Quai Ernest Ansermet, CH-1211 Geneva, Switzerland
3. National Centre for Computational Design and Discovery of Novel Materials (MARVEL), École Polytechnique Fédérale de Lausanne, CH-1015 Lausanne, Switzerland
4. National Institute for Materials Science, 1-1 Namiki, Tsukuba, 305-0044, Japan
5. These authors contributed equally.

*E-mail: Alberto.Morpurgo@unige.ch

**The magnetic state of atomically thin semiconducting layered antiferromagnets such as CrI$_3$ and CrCl$_3$ can be probed by forming tunnel barriers and measuring their resistance as a function of magnetic field ($H$) and temperature ($T$). This is possible because the tunneling magnetoresistance originates from a spin-filtering effect sensitive to the relative orientation of the magnetization in different layers, *i.e.*, to the magnetic state of the multilayers. For systems in which antiferromagnetism occurs within an individual layer, however, no spin-filtering occurs: it is unclear whether this strategy can work. To address this issue, we investigate tunnel transport through atomically thin crystals of MnPS$_3$, a van der Waals semiconductor that in the bulk exhibits easy-axis antiferromagnetic order within the layers. For thick multilayers below $T \sim 78$ K, a $T$-dependent magnetoresistance sets-in at $\mu_0 H \sim 5$ T, and is found to track the boundary between the antiferromagnetic and the spin-flop phases known from bulk magnetization measurements. The magnetoresistance persists down to individual MnPS$_3$ monolayers with nearly unchanged characteristic temperature and magnetic field scales, albeit with a different dependence on *H*. We discuss the implications of these finding for the magnetic state of atomically thin MnPS$_3$ crystals, conclude that antiferromagnetic correlations persist down to the level of individual** **monolayers, and that tunneling magnetoresistance does allow magnetism in 2D insulating materials to be detected even in the absence of spin-filtering.**

Probing the occurrence of magnetism in atomically thin crystals(*1-5*) is difficult because experimental techniques that are conventionally applied to bulk crystals (neutron diffraction(*6, 7*), magnetization measurements(*8, 9*), *etc.*) are not sufficiently sensitive to work at the atomic scale. Recent work on so-called layered antiferromagnets(*5, 10-21*) has shown that measuring the temperature-dependent magnetoresistance of tunnel barriers provides information about their magnetic state, and even allows their magnetic phase diagram to be determined(*22*). That is so because in these antiferromagnets the spins within each individual layer are ferromagnetically aligned, and their magnetic state is fully determined by the relative orientation of the layer magnetization vectors(*12, 15, 20, 21*). For magnetic configurations of this type, the sensitivity of the magnetoresistance to the magnetic states originates from a spin-filtering effect, such that aligning the magnetization of individual layers causes a reduction in resistance(*12-15, 20-25*). For van der Waals (vdW) compounds in which antiferromagnetism occurs within an individual layer, however, no such spin-filtering effect is present, the same logic does not apply. For atomically thin crystals –and certainly for monolayers– this situation is problematic, because if transport measurements cannot be used, it is not obvious what other technique could be employed to detect antiferromagnetism (for interesting attempts based on Raman spectroscopy see Ref. (*26, 27*)). To address this issue, we perform tunneling magnetoresistance measurements on atomically thin crystals of $MnPS_3$ and show that they allow tracking experimentally the phase boundary between the antiferromagnetic and the spin-flop phases of these 2D systems, all the way down to the ultimate limit of individual monolayers.

$MnPS_3$ is an exfoliable 2D material(*25, 28-30*) whose properties in bulk form have been investigated in the past. It is known that in bulk $MnPS_3$ crystals antiferromagnetism sets in at $T_N = 78$ K with the spins of the manganese atoms ordering within individual layers, and pointing nearly perpendicularly(*31*) to them due to an easy-axis out-of-plane anisotropy (Fig. 1(a))(*6*). As the antiferromagnetic exchange is much stronger than the anisotropy energy(*32*), upon the application of magnetic field perpendicular to the layers a spin-flop transition(*33, 34*) occurs at $H_{sf}$. In the spin-flop state the magnetic moments on the Mn atoms preserve their antiferromagnetic ordering, but re-align to point predominantly in the plane, with an out-of-plane component that increases upon increasing *H* (Ref. (*7*)). The spin-flop transition in bulk MnPS$_3$ crystals is easy to detect, as its occurrence is signaled by a well-defined onset of the out-of-plane magnetization *M* (red curve in Fig. 1(b)), and a concomitant peak in the differential magnetic susceptibility $\chi_m = \frac{1}{\mu_0}\frac{dM}{dH}$ centered at $\mu_0 H_{sf} = 5.3\ T$ (blue curve in Fig. 1(b)). At a more microscopic level, theoretical analyses of early experiments as well as more recent experimental work indicate that long-range magnetic dipolar interactions play an important role in stabilizing the antiferromagnetic state in MnPS$_3$, and in accounting for its properties(*7, 35-38*).

These characteristics make MnPS$_3$ an ideal candidate to explore whether it is possible to investigate the evolution of the magnetic state down to individual monolayers by means of tunneling magnetoresistance measurements. If dipolar interactions are as important as currently believed, the effect of fluctuations on the magnetic state is suppressed at low temperature and antiferromagnetism can persist in the two-dimensional limit, without a drastic reduction in transition temperature as compared to the bulk(*39*) (indeed, Mermin-Wagner theorem that precludes the existence of long-range order in 2D is valid for short ranged interactions and does not apply if long-ranged dipolar interactions play a relevant role). Additionally, the occurrence of the spin-flop transition at experimentally accessible values of applied field provides a hallmark signature of the antiferromagnetic state that we can search for in tunneling magnetoresistance measurements. Specifically, for MnPS$_3$ multilayers that are sufficiently thick we expect the signatures of the spin-flop transition to occur at the same temperature and magnetic field as in bulk crystals, enabling its identification. Once the manifestation of the spin-flop transition on the tunneling magnetoresistance has been identified, we can proceed to probe whether the phenomenon persists as the MnPS$_3$ thickness is decreased down to a single monolayer.

Following this strategy, we have performed magnetoresistance measurements on MnPS$_3$ tunnel barriers with thickness ranging from 13-layer down to an individual monolayer. The devices consist of an atomically thin MnPS$_3$ crystal contacted with few-layer graphene stripes in a cross geometry (see the inset of Fig. 1(c)), and encapsulated between two exfoliated boron nitride crystals (to prevent MnPS$_3$ degradation the devices are assembled in a glove box with controlled Nitrogen atmosphere, see Supplementary Information for details). The very weak temperature dependence of the resistance measured at low *T* –as contrasted to the thermally activated behavior observed at higher *T* (see Fig. 1c)– indicates that low-temperature transport does indeed occur in the tunneling regime(*40*). For barriers up to 4-layer thick, direct tunneling gives a sufficiently large current to be detected experimentally, resulting in linear *I-V* curves at low bias. For thicker barriers the probability for direct tunneling is too small and a large bias needs to be applied to generate a measurable current, resulting in strongly non-linear *I-V* curves (Fig. 1(d)). In this latter case, tunneling occurs in the so-called Fowler-Nordheim regime(*41, 42*), with $ln\frac{J}{V^2}$ scaling linearly with $\frac{1}{V}$, i.e. $ln\frac{J}{V^2} = -\frac{4t\sqrt{2m^*\phi_B^3}}{3\hbar e}\frac{N}{V} + C$ (see Fig. 1(e); here $J = I/S$, with $S$ the junction area, $N$ is the number of layers, $t$ = 0.65 nm the interlayer separation, $\phi_B$ is the barrier height, $m^*$ the effective mass of MnPS$_3$, and $e$ the electron charge). In agreement with expectations, the conductance in the linear regime decreases exponentially with increasing the multilayer thickness $d = N \times t$ (Fig. 1(f)), and in the Fowler-Nordheim regime the coefficient $\alpha$ of proportionality between $ln\frac{J}{V^2}$ and $\frac{1}{V}$ decreases linearly with $d$ (Fig. 1(g))(*41*). From the slopes of these two curves we can estimate the height of the tunnel barrier in few-layer graphene/MnPS$_3$ devices, as well as the effective mass $m^*$ in MnPS$_3$, which we find to be respectively $\phi_B \cong 440$ meV and $m^* \approx 0.5\, m_0$ (with $m_0$ the free-electron mass).

We start by searching for the presence of tunneling magnetoresistance $\eta(H) = \frac{R(H)-R(0)}{R(0)}$ in relatively thick multilayers, to establish whether the occurrence of the spin-flop transition does give origin to a detectable signal. To this end, we measure the temperature and magnetic field dependence of the resistance of tunnel barriers realized using $N = 13$ (13L) and $N = 6$ (6L) $MnPS_3$ multilayers (Fig. 2(a) and 2(b)). For both devices at low temperature, the resistance is independent of applied field $H$ up to a threshold $\mu_0 H_1 \cong 5$ T —very close to the magnetic field value at which the spin-flop transition is seen in bulk crystals (Fig. 1(b)). Above this threshold a pronounced increase takes place and, as the applied field is increased further, a change in slope of the magnetoresistance occurs at $\mu_0 H_2 \cong 7.5$ T. For both 13L and 6L $MnPS_3$ multilayers, the total low-temperature magnetoresistance that we observe at the highest field reached in our experiments ($\mu_0 H \cong 12$ T) is approximately 15 %, much smaller than in tunnel barriers of layered antiferromagnetic insulators(*12, 15, 20-22*), but still sizable and easily measurable.

Upon increasing temperature $T$, the magnitude of the magnetoresistance decreases and eventually disappears (becomes comparable to the noise) for $T$ between 70 and 80 K (Fig. 2(a) and (b)). In order to identify more precisely the value of $T$ at which the resistance stops depending on magnetic field, we plot the magnetoresistance at a fixed value of $H$ as a function of temperature, and extrapolate the data points to determine the value of $T$ for which the magnetoresistance vanishes; see Fig. 2(c) for the 13L device and Fig. 2(d) for the 6L device. For the 13L and 6L devices we find that the magnetoresistance vanishes for temperature higher than $76\pm5$ K and $78\pm3$ K respectively, very close to the Néel temperature $T_N = 78$ K of bulk $MnPS_3$ crystals(*43*). Within the experimental precision, the extracted values do not depend on the specific value of $H$ used to perform the analysis, see Fig. 2(c) and 2(d). We emphasize, nevertheless, that care is needed in comparing the observed onset temperature to the Néel temperature, because the observation of any magnetoresistance requires a field $\mu_0 H > 5$ T to be applied whereas the Néel temperature is defined as the transition from the paramagnetic to the antiferromagnetic state at zero applied field. Indeed, as we make clear below, the onset temperature for the appearance of magnetoresistance corresponds to the critical temperature for the transition between the spin-flop and the paramagnetic state.

The measurements just discussed show that the 13L and the 6L devices exhibit a virtually identical behavior, in agreement with the notion that their thickness is sufficiently large to approach the magnetic behavior of bulk $MnPS_3$ crystals. Consistently with this conclusion, all observations appear to indicate that the characteristic magnetic and temperature scales detected in the tunneling magnetoresistance coincide with those at which magnetic phase transitions take place in bulk crystals. To establish the nature of the relation between the magnetoresistance of atomically thin multilayers and magnetic phase boundaries in bulk $MnPS_3$ more precisely, we represent the magnetoresistance measured on the 6L device as a color plot (Fig. 3(a)). Onto the same plot we overlay the spin-flop field $H_{sf}$ as a function of temperature (orange diamond) and the Néel temperature $T_N$ as a function of magnetic field (orange circle), obtained by Goossens *et al.* in their systematic study of $MnPS_3$ bulk magnetization(*9*). The orange diamonds outline the boundary between the antiferromagnetic phase stable at low field and the spin-flop phase that is stable at higher field: it is apparent from Fig. 3(a) that this phase boundary coincides with the observed onset of the magnetoresistance in the 6L tunnel barrier (note how the temperature for the transition between the paramagnetic and antiferromagnetic states is nearly independent of $H$. This is why the onset temperature for the occurrence of magnetoresistance at $\mu_0 H > 5$ T corresponds well to the value of $T_N$ in the bulk).

The results of these measurements already allow us to draw two important conclusions. First, a sizable tunneling magnetoresistance is present in $MnPS_3$ despite the absence of any spin-filtering effect. Without discussing the microscopic mechanism responsible for the measured magnetoresistance in $MnPS_3$ (see below), we emphasize that a difference with previously investigated layered antiferromagnets is clear, since in $MnPS_3$ the magnetoresistance is positive whereas in $CrI_3$ and $CrCl_3$ it is negative. The second important conclusion is that the analysis of the temperature and magnetic field dependence of the tunneling resistance of the 13L and 6L $MnPS_3$ devices allows the presence of the magnetic transition from the antiferromagnetically ordered state to the spin-flop state to be unambiguously detected (see Fig. 3(a)). It is this sensitivity of the tunneling magnetoresistance to the phase boundaries that provides useful information about the magnetic state of the barrier material.

Interestingly, magnetoresistance measurements appear to be more sensitive to details of the magnetic state than measurements done on bulk crystals. This is illustrated by the observation of two distinct characteristic fields $H_1$ and $H_2$ –having virtually identical values in both the 13L and 6L devices, see Fig. 2(a) and 2(b)– and not of a single spin-flop field. The "splitting" of the spin-flop transition has been predicted theoretically for antiferromagnets whose magnetic anisotropy originates from both on-site and inter-site anisotropy(*44-49*), since in the appropriate regime the competition between these anisotropy terms makes the magnetic state evolve through a sequence of distinct steps as $H$ is swept from below $H_1$ to above $H_2$. In simple terms, the transition from the antiferromagnetic to the spin-flop state occurs through intermediate spin configurations, such that for $H_1 < H < H_2$ the antiferromagnetically coupled spins are oriented along a direction that evolves gradually from being parallel (at $H = H_1$) to perpendicular (at $H = H_2$) to the applied field (details depend on the precise values of microscopic parameters). In experiments on bulk crystals, the competition of different anisotropy terms has indeed been invoked to interpret neutron scattering experiments on MnPS$_3$ (Ref. (*50*)), but a splitting of the spin-flop transition was never reported in magnetization measurements because inhomogeneity and disorder (such as domain walls(*33*)) cause a large broadening of the corresponding peak in the magnetic susceptibility (Fig. 1(b)). Owing to their very small size, tunnel barrier devices are much less influenced by inhomogeneity of structural and magnetic origin as compared to bulk crystals, enabling subtler features to be detected. Under these conditions, we compare the characteristic fields observed in the 13L and 6L with the bulk spin-flop field $H_{sf}$, by defining –inspired by theory(*44*)– an "effective" spin-flop field $H_{sf}^e = \sqrt{H_1 \times H_2}$. Fig. 3(b) shows that $H_{sf}^e$ matches well $H_{sf}$ throughout the temperature range investigated, which strengthens our confidence in the proposed interpretation of the fields $H_1$ and $H_2$.

Irrespective of these details, having observed tunneling magnetoresistance in thick MnPS$_3$ multilayers and identified the presence of a spin-flop transition as a distinctive signature of antiferromagnetism, we are now in the position to extend our experiments to the ultimate limit of individual monolayer barriers. The results of measurements performed on mono- and bilayer tunnel barriers –shown in Fig. 4(a) and 4(b)– exhibit important similarities as well as clear differences to the behavior observed in the 13L and 6L devices. Notable differences are the functional dependence of the magnetoresistance, which starts varying already for small applied magnetic field (without a threshold as for thick multilayers), and its larger magnitude (in monolayer, the magnetoresistance reaches values close to 100%, as compared to 15% observed in the 13- and 6-layer devices). The similarities are more telling. The magnetoresistance exhibits a characteristic magnetic field scale –the position of the pronounced peak observed in Fig. 4(a) and 4(b)– that corresponds well to the spin-flop field measured in thicker multilayers: the peak occurs at $\mu_0 H \cong 4$ T in monolayers and at $\mu_0 H \cong 5$ T in bilayers (close to $\mu_0 H_{sf} = 5.3$ T, the spin-flop field in bulk MnPS$_3$). Even more strikingly, the magnetoresistance starts depending on temperature as $T$ is lowered below approximately 75 K (respectively 78±5 K and 74 ± 5 K, see Fig. 4(c) and 4(d)), i.e. nearly exactly the temperature values observed in the 13L and 6L devices. These findings are summarized in Fig. 4(e) and (f). Fig. 4(e) shows the evolution of the characteristic field extracted from the magnetoresistance measurements as a function of multilayer thickness (for 13L and 6L the plotted value corresponds to $H_{sf}^e = \sqrt{H_1 \times H_2}$, see discussion above), and compares it with the spin-flop field of bulk MnPS$_3$ (represented by the dashed horizontal line). Fig. 4(f) does the same thing for the characteristic temperature that, within the experimental uncertainty, is essentially independent of thickness and coincides with the bulk Néel temperature.

Finding that a pronounced tunneling magnetoresistance continues to be observed as the MnPS$_3$ thickness is reduced allows us to conclude –directly from the experimental data– that magnetism persists in mono and bilayer MnPS$_3$ (simply because non-magnetic tunnel barriers such as hBN(*51, 52*) exhibit no tunneling magnetoresistance). This is a non-trivial conclusion, because it has been very recently suggested on the basis of temperature-dependent Raman measurements that monolayer MnPS$_3$ is non-magnetic(*27*). To extract information about the nature of the magnetic state in mono and bilayer MnPS$_3$, we start by discussing the aspects of the measured magnetoresistance that are common to all layers, irrespective of their thickness. In particular, for sufficiently thick MnPS$_3$ multilayers (i.e., the 13L and 6L devices) the *T*- and *H*-dependence of the magnetoresistance onset precisely tracks the temperature evolution of the spin-flop field (see Fig. 3(a)), which allows us to conclude with confidence that in these multilayers the magnetic state coincides with that of bulk crystals (i.e., it is an antiferromagnetic state at low applied field, for $\mu_0 H < 5$T). In mono and bilayers the dependence of the magnetoresistance on *H* differs from that observed in the 13L and 6L devices –indicating that the magnetic state is also different– but the magnetic field and temperature scale extracted from the experiments remain nearly unaltered (Fig. 4(e) and 4(f)). From this observation we infer that a spin-flop transition persists down to the ultimate thickness of individual monolayers and –since such a transition is characteristically associated to weakly anisotropic antiferromagnetic ordering– we conclude that antiferromagnetic correlations remain present even in the ultimate limit of individual monolayers.

These conclusions are robust because they rely on qualitative features directly visible in the measured tunneling magnetoresistance, which allow boundaries of the magnetic phase diagram to be tracked. Establishing more quantitative information about details of the magnetic states is however difficult, because the only experimentally accessible physical quantity in our measurements is the dependence of the tunneling resistance on magnetic field and temperature, for which there currently exists no well-defined microscopic model to perform a quantitative analysis. For instance, the statement that antiferromagnetic correlations persist all the way down to individual monolayers does not necessarily imply that these correlations are long-ranged. As we discussed in the introduction, owing to the presence of dipolar interactions in MnPS$_3$, long-range antiferromagnetic order can persist in monolayers, but to generate tunneling magnetoresistance a short correlation length is likely sufficient (on physical grounds, magnetic correlations over a length comparable to the Fermi wavelength in the tunneling electrodes are probably enough). For the same reason, we cannot conclude whether the different magnetic field dependence of the resistance observed in mono and bilayers as compared to 13L and 6L devices originates from differences in the ground state spin-configuration (certainly expected, since –in the presence of long-range dipolar interactions– reducing the number of layers also reduces the energy balance determining the spin configuration) or from enhanced spin-fluctuations (always present upon reducing dimensionality).

Detailed microscopic models to describe tunneling transport through magnetic barriers will need to be developed to answer this type of questions. At this stage we can only try to identify physical mechanisms determining the magnetoresistance of magnetic tunnel barriers. An obvious one is that, in the presence of spin-orbit interaction, a change in magnetic state leads to a modification in the band structure. In a semiconducting material such as $MnPS_3$ this can cause a change in the size of the band gap, and therefore in the height of the tunnel barrier $\phi_B$ in our devices. Indeed, the $MnPS_3$ band structure obtained from *ab-initio* calculations shows clear differences in the antiferromagnetic and in the spin-flop states (see respectively the blue and red lines in Fig. 3(c)), with the degeneracy between K and K' valleys that is lifted in the antiferromagnetic state(*53*) and recovered in the spin-flop phase. As a result of these differences, the band gap of $MnPS_3$ is slightly larger (by an amount of the order of 10 meV) in the spin-flop phase, in accord with the observed increase in resistance as $H > H_{sf}$. Albeit possibly only coincidental –because the precision of *ab-initio* methods to determine quantitatively semiconducting band gaps is limited– the calculations capture the correct order of magnitude of the effect: with the height of the tunnel barrier estimated from the analysis of tunneling process, $\phi_B \approx 440$ meV, the 15% magnetoresistance observed in the 13L and 6L devices requires a variation $\Delta\phi_B \cong 10$ meV. Extending this analysis beyond an order of magnitude estimate is however complex, because detailed calculations of the *H*-dependent tunneling resistance require as input the spin-configuration as a function of applied field.

It is clear from these considerations that a great deal of theoretical work will be needed to analyze the details of the electron tunneling processes in antiferromagnetic tunnel barriers, and to extract microscopic information about the magnetic state of atomically thin $MnPS_3$ layers from the field dependence of the tunneling magnetoresistance (*e.g.*, information about spin fluctuations, correlation length, *etc.*). Nevertheless, even just the fact that tunneling magnetoresistance measurements enable the persistence of antiferromagnetic correlations to be confirmed down to individual monolayers is an important finding, since it is very far from obvious which other experimental routes could be followed to establish such a conclusion. For ferromagnetic monolayers, Kerr effect(*5*) or direct magnetometry(*54*) have been recently demonstrated as possible techniques, but these same techniques would not work in the absence of a net magnetization. Possibly, spin polarized scanning tunneling microscopy(*55*) could be an alternative, but the technical complexities of performing such experiments on monolayer antiferromagnets are considerably higher (even more so if we consider the demands that this technique poses on sample preparation). The observations reported here, therefore, are not only relevant because they reveal the persistence of antiferromagnetic correlations in the truly 2D limit of individual monolayers, but also because they open a new experimental route that can be applied to a broad class of 2D antiferromagnetic materials.

**Acknowledgements**

The authors gratefully acknowledge A. Ferreira for technical support, A. Waelchli and S. Gariglio for their assistance with the measurements. A.F.M. acknowledges financial support from the Swiss National Science Foundation (Division II) and from the EU Graphene Flagship project. M.G. acknowledges support from the Swiss National Science Foundation through the Ambizione program. Growth of hexagonal boron nitride crystals was supported by the Elemental Strategy Initiative conducted by the MEXT, Japan, A3 Foresight by JSPS and the CREST(JPMJCR15F3), JST.

**Author contributions**

G.L. and H.H. fabricated and measured all the devices with the help of Z.W. and analyzed the data under the supervision of A.F.M. Together with M.G., who performed the *ab-initio* calculations and A.F.M., G.L. and H.H. interpreted the results. D.D. and E.G. grew and characterized the MnPS$_3$ crystals. T.T. and K.W. provided high quality boron nitride crystals. All authors contributed to writing the manuscript.

**Competing interests**

The authors declare no competing interests.

**Data availability**

The data related to this study are available from the corresponding author upon reasonable and well-motivated request.

**Code availability**

The code related to this study are available from the corresponding author upon reasonable and well-motivated request.

**References**

1. K. S. Burch, D. Mandrus, J.-G. Park, Magnetism in two-dimensional van der Waals materials. *Nature* **563**, 47 (2018/11/01, 2018).
2. C. Gong, X. Zhang, Two-Dimensional Magnetic Crystals and Emergent Heterostructure Devices. *Science* **363**, 706 (2019).
3. M. Gibertini, M. Koperski, A. F. Morpurgo, K. S. Novoselov, Magnetic 2D materials and heterostructures. *Nature Nanotechnology* **14**, 408 (2019/05/01, 2019).
4. C. Gong *et al.*, Discovery of intrinsic ferromagnetism in two-dimensional van der Waals crystals. *Nature* **546**, 265 (2017).
5. B. Huang *et al.*, Layer-dependent ferromagnetism in a van der Waals crystal down to the monolayer limit. *Nature* **546**, 270 (2017).
6. K. Kurosawa, S. Saito, Y. Yamaguchi, Neutron diffraction study on mnps3 and feps3. *Journal of the Physical Society of Japan* **52**, 3919 (1983).

7. D. Goossens, A. Wildes, C. Ritter, T. Hicks, Ordering and the nature of the spin flop phase transition in MnPS3. *Journal of Physics: Condensed Matter* **12**, 1845 (2000).
8. K. Okuda *et al.*, Magnetic properties of layered compound mnps3. *Journal of the Physical Society of Japan* **55**, 4456 (1986).
9. D. J. Goossens, T. J. Hicks, The magnetic phase diagram of $Mn_xZn_{1-x}PS_3$. *Journal of Physics: Condensed Matter* **10**, 7643 (1998, 1998).
10. B. Huang *et al.*, Electrical control of 2D magnetism in bilayer CrI 3. *Nature nanotechnology* **13**, 544 (2018).
11. S. Jiang, L. Li, Z. Wang, K. F. Mak, J. Shan, Controlling magnetism in 2D CrI 3 by electrostatic doping. *Nature nanotechnology* **13**, 549 (2018).
12. Z. Wang *et al.*, Very large tunneling magnetoresistance in layered magnetic semiconductor CrI 3. *Nature communications* **9**, 2516 (2018).
13. X. Cai *et al.*, Atomically Thin CrCl3: An in-Plane Layered Antiferromagnetic Insulator. *Nano letters*, (2019).
14. D. R. Klein *et al.*, Giant enhancement of interlayer exchange in an ultrathin 2D magnet. Preprint at https://arxiv.org/abs/1903.00002 (2019, 2019).
15. H. H. Kim *et al.*, One Million Percent Tunnel Magnetoresistance in a Magnetic van der Waals Heterostructure. *Nano Letters* **18**, 4885 (2018/08/08, 2018).
16. S. Jiang, J. Shan, K. F. Mak, Electric-field switching of two-dimensional van der Waals magnets. *Nature materials* **17**, 406 (2018).
17. H. H. Kim *et al.*, Evolution of interlayer and intralayer magnetism in three atomically thin chromium trihalides. *Proceedings of the National Academy of Sciences* **116**, 11131 (2019).
18. S. Jiang, L. Li, Z. Wang, J. Shan, K. F. Mak, Spin tunnel field-effect transistors based on two-dimensional van der Waals heterostructures. *Nature Electronics* **2**, 159 (2019/04/01, 2019).
19. T. Song *et al.*, Voltage Control of a van der Waals Spin-Filter Magnetic Tunnel Junction. *Nano Letters* **19**, 915 (2019/02/13, 2019).
20. T. Song *et al.*, Giant tunneling magnetoresistance in spin-filter van der Waals heterostructures. *Science* **360**, 1214 (2018).
21. D. R. Klein *et al.*, Probing magnetism in 2D van der Waals crystalline insulators via electron tunneling. *Science* **360**, 1218 (2018).
22. Z. Wang *et al.*, Determining the Phase Diagram of Atomically Thin Layered Antiferromagnet $CrCl_3$. *Submitted*.
23. Z. Wang *et al.*, Tunneling spin valves based on Fe3GeTe2/hBN/Fe3GeTe2 van der Waals heterostructures. *Nano letters* **18**, 4303 (2018).
24. T. R. Paudel, E. Y. Tsymbal, Spin Filtering in CrI3 Tunnel Junctions. *ACS Applied Materials & Interfaces* **11**, 15781 (2019/05/01, 2019).
25. H. H. Kim *et al.*, Tailored Tunnel Magnetoresistance Response in Three Ultrathin Chromium Trihalides. *Nano Letters* **19**, 5739 (2019).
26. J.-U. Lee *et al.*, Ising-type magnetic ordering in atomically thin FePS3. *Nano letters* **16**, 7433 (2016).
27. K. Kim *et al.*, Antiferromagnetic ordering in van der Waals 2D magnetic material MnPS3 probed by Raman spectroscopy. *2D Materials* **6**, 041001 (2019/07/12, 2019).
28. G. Long *et al.*, Isolation and Characterization of Few-Layer Manganese Thiophosphite. *ACS nano* **11**, 11330 (2017).
29. W. Xing *et al.*, Magnon transport in quasi-two-dimensional van der Waals antiferromagnets. *Physical Review X* **9**, 011026 (2019).

30. K.-z. Du *et al.*, Weak van der Waals stacking, wide-range band gap, and Raman study on ultrathin layers of metal phosphorus trichalcogenides. *ACS nano* **10**, 1738 (2015).
31. E. Ressouche *et al.*, Magnetoelectric MnPS 3 as a candidate for ferrotoroidicity. *Physical Review B* **82**, 100408 (2010).
32. A. Wildes, B. Roessli, B. Lebech, K. Godfrey, Spin waves and the critical behaviour of the magnetization in MnPS3. *Journal of Physics: Condensed Matter* **10**, 6417 (1998).
33. L. Néel, Propriétés magnétiques de l'état métallique et énergie d'interaction entre atomes magnétiques. *Ann. Phys.* **11**, 232 (1936).
34. H. De Groot, L. De Jongh, Phase diagrams of weakly anisotropic Heisenberg antiferromagnets, nonlinear excitations (solitons) and random-field effects. *Physica B+ C* **141**, 1 (1986).
35. T. Hicks, T. Keller, A. Wildes, Magnetic dipole splitting of magnon bands in a two dimensional antiferromagnet. *Journal of Magnetism and Magnetic Materials* **474**, 512 (2019).
36. D. Goossens, Dipolar anisotropy in quasi-2D honeycomb antiferromagnet MnPS 3. *The European Physical Journal B* **78**, 305 (2010).
37. C. Pich, F. Schwabl, Spin-wave dynamics of two-dimensional isotropic dipolar honeycomb antiferromagnets. *Journal of magnetism and magnetic materials* **148**, 30 (1995).
38. A. Wildes, H. Rønnow, B. Roessli, M. Harris, K. Godfrey, Anisotropy and the critical behaviour of the quasi-2D antiferromagnet, MnPS3. *Journal of magnetism and magnetic materials* **310**, 1221 (2007).
39. C. Pich, F. Schwabl, Order of two-dimensional isotropic dipolar antiferromagnets. *Physical Review B* **47**, 7957 (1993).
40. S. M. Sze, K. K. Ng, *Physics of semiconductor devices*. (John wiley & sons, 2006).
41. J. G. Simmons, Generalized formula for the electric tunnel effect between similar electrodes separated by a thin insulating film. *Journal of applied physics* **34**, 1793 (1963).
42. M. Lenzlinger, E. Snow, Fowler‐Nordheim tunneling into thermally grown SiO2. *Journal of Applied physics* **40**, 278 (1969).
43. P. Joy, S. Vasudevan, Magnetism in the layered transition-metal thiophosphates M PS 3 (M= Mn, Fe, and Ni). *Physical Review B* **46**, 5425 (1992).
44. B. Ivanov, Mesoscopic antiferromagnets: statics, dynamics, and quantum tunneling. *Low temperature physics* **31**, 635 (2005).
45. N. Yamashita, Field Induced Phase Transitions in Uniaxial Antiferromagnets. *Journal of the Physical Society of Japan* **32**, 610 (1972/03/15, 1972).
46. K.-S. Liu, M. E. Fisher, Quantum lattice gas and the existence of a supersolid. *Journal of Low Temperature Physics* **10**, 655 (1973/03/01, 1973).
47. C. C. Becerra, L. G. Ferreira, Phase Transitions in Uniaxial Antiferromagnets. *Journal of the Physical Society of Japan* **37**, 951 (1974/10/15, 1974).
48. H.-F. Li, Possible ground states and parallel magnetic-field-driven phase transitions of collinear antiferromagnets. *Npj Computational Materials* **2**, 16032 (10/14/online, 2016).
49. J. A. J. Basten, W. J. M. de Jonge, E. Frikkee, Intermediate phase in the spin-flop system Co${\mathrm{Br}}_{2}$ \ifmmode\cdot\else\textperiodcentered\fi{} 6(0.48${\mathrm{D}}_{2}$O, 0.52${\mathrm{H}}_{2}$O). *Physical Review B* **21**, 4090 (05/01/, 1980).
50. A. Wildes, H. Rønnow, B. Roessli, M. Harris, K. Godfrey, Static and dynamic critical properties of the quasi-two-dimensional antiferromagnet MnPS 3. *Physical Review B* **74**, 094422 (2006).

51. G.-H. Lee *et al.*, Electron tunneling through atomically flat and ultrathin hexagonal boron nitride. *Applied Physics Letters* **99**, 243114 (2011/12/12, 2011).
52. L. Britnell *et al.*, Electron Tunneling through Ultrathin Boron Nitride Crystalline Barriers. *Nano Letters* **12**, 1707 (2012/03/14, 2012).
53. X. Li, T. Cao, Q. Niu, J. Shi, J. Feng, Coupling the valley degree of freedom to antiferromagnetic order. *Proceedings of the National Academy of Sciences*, 201219420 (2013).
54. L. Thiel *et al.*, Probing Magnetism in 2D Materials at the Nanoscale with Single-Spin Microscopy. *Science* **364**, 973 (2019).
55. A. Kubetzka *et al.*, Revealing Antiferromagnetic Order of the Fe Monolayer on W(001): Spin-Polarized Scanning Tunneling Microscopy and First-Principles Calculations. *Physical Review Letters* **94**, 087204 (03/03/, 2005).

**Figures**

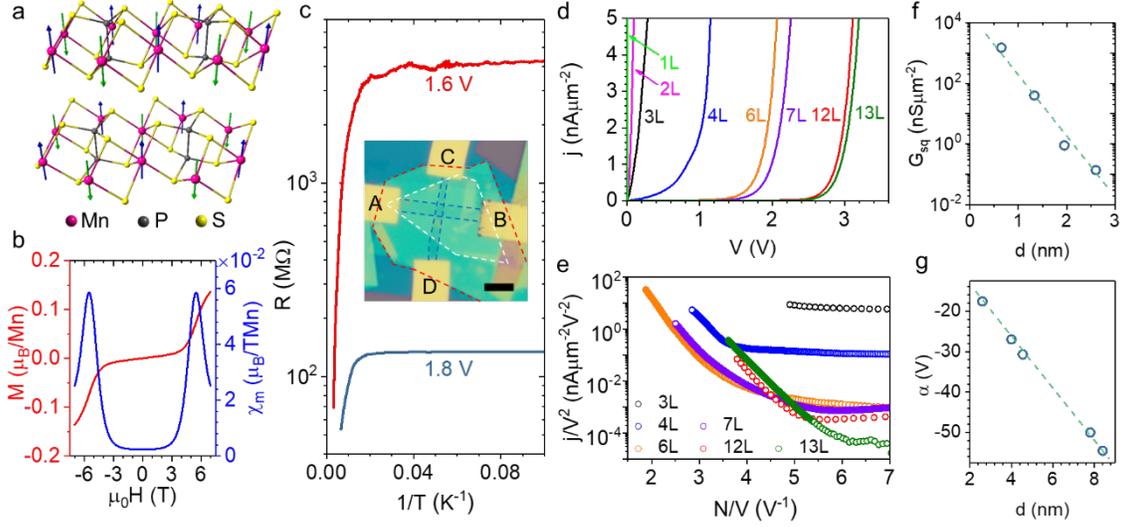

**Figure 1. MnPS$_3$ tunnel junctions.** (a) MnPS$_3$ crystal structure, illustrating the antiferromagnetically aligned spins on the Mn atoms (note: past neutron studies have shown that the spin direction is not exactly perpendicular to the layers). (b) Magnetization (red curve) and differential magnetic susceptibility (blue curve) of bulk MnPS$_3$ measured at $T = 10$ K as a function of perpendicular magnetic field. A spin-flop transition around $\mu_0 H \cong 5$ T manifests itself in the increase of magnetization accompanied by a peak in differential susceptibility. (c) Resistance of a 6-layer MnPS$_3$ device measured with 1.6 V (red) and 1.8 V (blue) applied bias, showing a thermally activated behavior at high $T$ followed by near saturation at low $T$, indicative of tunneling transport. The inset shows an optical microscope image of a device (the nano-fabricated gold pads, labelled with A, B, C, and D, contact the graphene stripes acting as electrodes; the white, blue and red dashed lines outline respectively the MnPS$_3$, graphene, and top hBN layer. The blue background is the bottom hBN flake. The bar is 5 μm). (d) Tunneling current density $j$ measured at $T =$ 1.5 K as a function of bias for MnPS$_3$ barriers of varying thicknesses (see the legend). (e) At high bias, in the Fowler-Nordheim tunneling regime, $\ln \frac{j}{V^2}$ is proportional to $\frac{N}{V}$ (with $N$ the number of MnPS$_3$ layers). (f) The linear tunneling conductance (per unit area) $G_{sq} = j/V$ decreases exponentially with MnPS$_3$ thickness $d = N \times t$ (here $t = 0.65$ nm is the interlayer separation), and (g) the coefficient $\alpha$ of proportionality between $\ln \frac{j}{V^2}$ and $\frac{N}{V}$ decreases linearly with $d$, as expected (the dashed green lines are linear fits to the data).

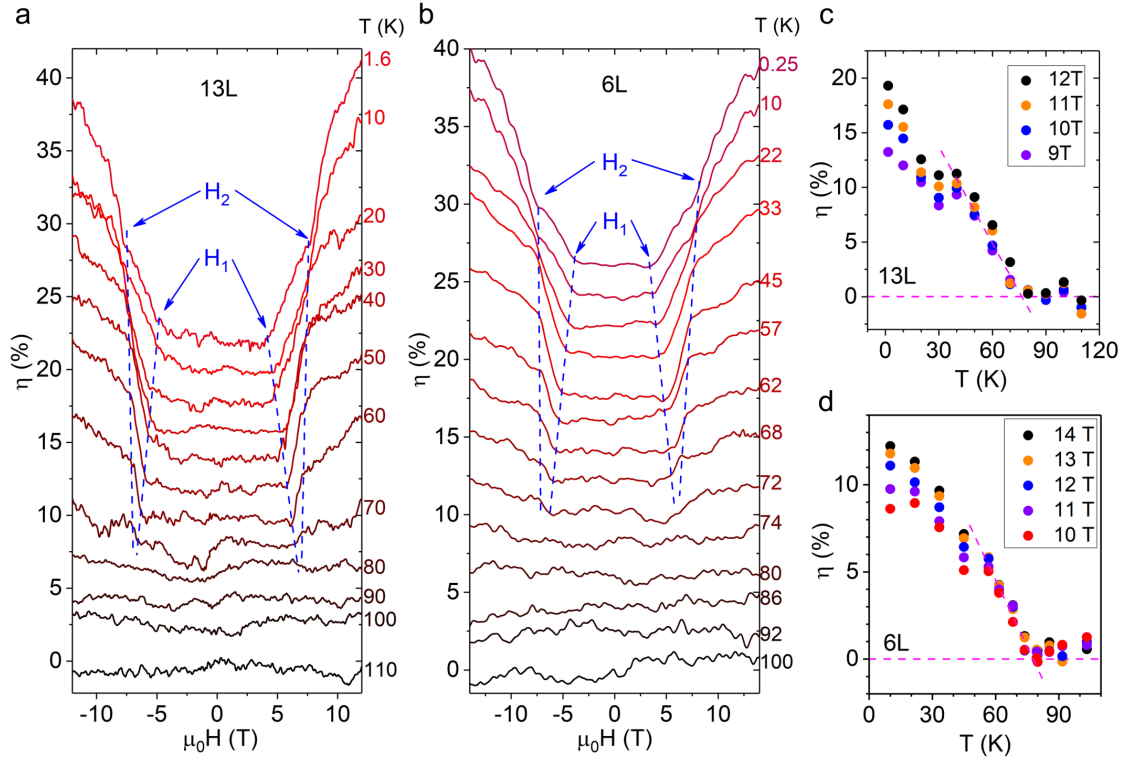

**Figure 2. Tunneling magnetoresistance of thick MnPS$_3$ multilayers.** Tunneling magnetoresistance $\eta\,(H) = \frac{R(H)-R(0)}{R(0)}$ measured at different values of temperature $T$ (as indicated in the respective panels) on a 13-layer (a) and a 6-layer (b) MnPS$_3$ tunnel barrier devices, showing virtually identical behavior (in both panels data are offset for clarity; at all temperatures $\eta\,(H=0)=0$ ). At low temperature, the tunneling resistance is independent of field up to approximately $\mu_0 H_1 \cong 5$ T, and increases past this threshold with a change in slope at $\mu_0 H_2 \cong 7.5$ T (see the blue arrows). Upon raising $T$, the magnetoresistance onset $H_1$ shifts to higher values and the magnetoresistance magnitude decreases; the field $H_2$ at which the slope of the magnetoresistance changes is approximately temperature independent (see the dashed blue lines). The magnetoresistance vanishes as $T$ is increased past 75 ~ 80 K, as determined by extrapolating the magnetoresistance measured at a fixed $H$ as a function of T (see panels (c) and (d) for the 13- and 6-layer device respectively).

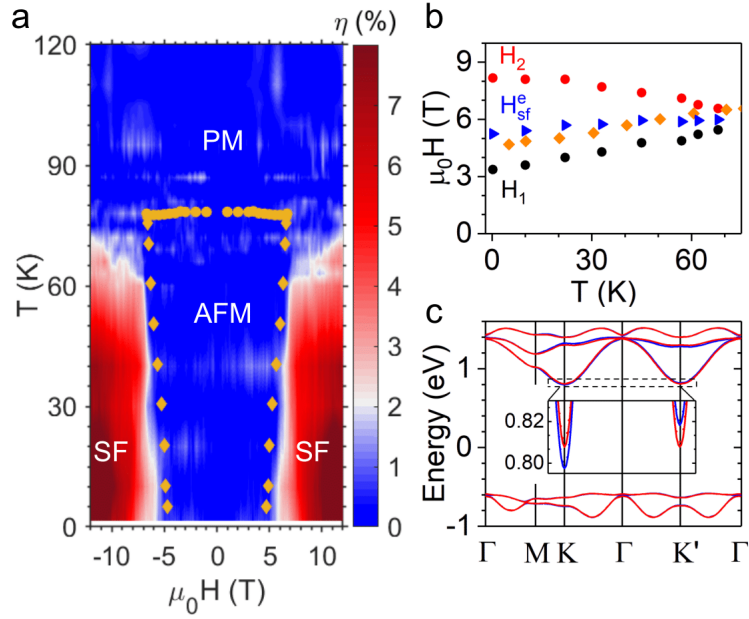

**Figure 3. Spin-flop transition probed by tunneling magnetoresistance in thick MnPS$_3$ multilayers.** (a) Color plot of the tunneling magnetoresistance $\eta$ (H,T) of 6-layer MnPS$_3$, as a function of applied magnetic field (from -12 to + 12 T) and temperature (from 1.5 to 120 K). The values of the spin-flop field $H_{sf}$ as a function of $T$ (orange diamonds) and of the Néel temperature $T_N$ as a function of $\mu_0 H$ (orange circles) extracted from past measurements on bulk MnPS$_3$ crystals (Ref. (*9*)) are overlaid on the same plot: note how the onset of magnetoresistance as a function of $\mu_0 H$ and $T$ coincide with the $T$-dependent spin-flop field $H_{sf}(T)$. (b) Temperature dependence of the characteristic fields $H_1$ (black circles) and $H_2$ (red circles) measured on a 6-layer tunnel barrier (see Fig. 2(b)), plotted together with the effective spin-flop field $H_{sf}^e = \sqrt{H_1 \times H_2}$ (blue triangles; see main text) and with the spin-flop field $H_{sf}$ measured on bulk crystals as reported in Ref. (*9*) (orange diamonds). It is apparent that $H_{sf}$ and $H_{sf}^e$ nearly coincide throughout the temperature range investigated. (c) Electronic band structure of MnPS$_3$ monolayers obtained from *ab-initio* calculations performed assuming the system to be in the antiferromagnetic phase (blue curves) or in the spin-flop phase (red curves). The inset zooms in at the bottom of the conduction band, showing that in the antiferromagnetic state the bottom of the conduction band near the K point is lower in energy, resulting in a slightly smaller energy gap and, accordingly, in a lower height of the MnPS$_3$ tunnel barrier.

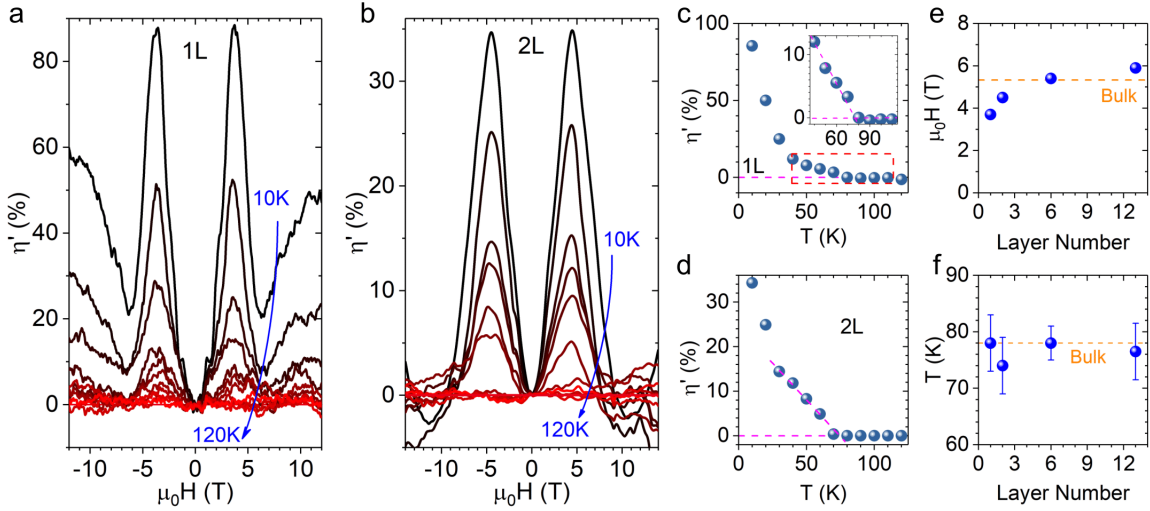

**Figure 4. Persistence of antiferromagnetic correlations in mono- and bilayer MnPS$_3$.** Tunneling magnetoresistance $\eta'$ ($H$) of monolayer (a) and bilayer (b) MnPS$_3$ as a function of applied field, measured as $T$ is increased from 10 to 120 K in 10 K steps. A small $T$-independent magnetoresistance persists up to the highest temperature of our measurements, likely due to the graphene electrodes (mono and bilayer devices have much smaller resistance than thicker tunnel barriers and the contact magnetoresistance is not entirely negligible; see also supplementary information), which is why we plot $\eta'$ ($H,T$) $\equiv \eta(H,T) - \eta(H, T = 120\ K)$. $\eta'$ ($H,T$) starts deviating from zero as $T$ is lowered below approximately 75 ~ 80 K (comparable to what is found in 13L and 6L devices) and increases upon cooling. In mono and bilayer devices $\eta'$ ($H,T$) exhibits no threshold at low field and peaks respectively at $\mu_0 H \cong 4$ T and $\mu_0 H \cong 5$ T, just slightly smaller than the bulk spin-flop field. Panels (c) and (d) show the peak magnetoresistance as a function of $T$, and its extrapolation to the value of $T$ for which $\eta'$ vanishes (78 and 74 K for mono and bilayer, respectively, in both cases with an error of approximately 5 K; the inset in (c) zooms in on the region close to the critical temperature). Panel (e) and (f) show the characteristic magnetic fields (the position of the peak in mono and bilayers, and the value of $H_{sf}^e$ in 13- and 6-layer devices), and the temperature at which magnetoresistance vanishes for all devices, as a function of layer number. The horizontal orange dashed line in the two panels indicate the bulk Néel temperature $T_N$ and the spin-flop field in bulk MnPS$_3$, respectively

Supplementary information for

# Spin-flop transition in atomically thin MnPS$_3$ crystals

Gen Long[1,2,5], Hugo Henck[1,2,5], Marco Gibertini[1,3], Dumitru Dumcenco[1], Zhe Wang[1,2], Takashi Taniguchi[4], Kenji Watanabe[4], Enrico Giannini[1], Alberto F. Morpurgo[1,2*]

1. Department of Quantum Matter Physics, University of Geneva, 24 Quai Ernest Ansermet, CH-1211 Geneva, Switzerland
2. Group of Applied Physics, University of Geneva, 24 Quai Ernest Ansermet, CH-1211 Geneva, Switzerland
3. National Centre for Computational Design and Discovery of Novel Materials (MARVEL), École Polytechnique Fédérale de Lausanne, CH-1015 Lausanne, Switzerland
4. National Institute for Materials Science, 1-1 Namiki, Tsukuba, 305-0044, Japan
5. These authors contributed equally.

*E-mail: Alberto.Morpurgo@unige.ch

## Table of contents

In the Method Section and in the supplementary notes here after we discuss a number of more technical points that could not be discussed in full detail in the main text. In particular, we present the details of the growth of MnPS$_3$ crystals and the basic aspects of their characterization (structural and magnetic), we outline the key points of the device fabrication process (including the determination of the multilayer thickness), and we show the temperature-independent background magnetoresistance that is present in tunnel barrier devices done using mono- and bilayer MnPS$_3$. Finally, we explain the technical details of the *ab initio* calculation of the band structure shown in Fig. 3(c) of the main text.

1. **Supplementary methods**

    1.1. **Crystal growth**

MnPS$_3$ crystals were grown by the Chemical Vapor Transport (CVT) method. A stoichiometric amount 1:1:3 of elements Mn (99.98%, Alfa Aesar), P (99.999%, Alfa Aesar), and Se (99.999%, Alfa Aesar), corresponding to approximately 1 gram total mass, were inserted in a quartz tube (inner diameter 10 mm) inside a glove box, together with approximately 2 mg/cm$^3$ of iodine (intended to act as transport agent during the growth process). The tube was subsequently evacuated down to $\sim 10^{-4}$ mbar, with intermediate Ar flushing, and sealed to a length of 16 cm. The sealed tube was inserted in a horizontal furnace, in a controlled temperature gradient of approximately 3 K/cm (with $T_{hot}$ = 680 °C and $T_{cold}$ = 630 °C, for the hot and cold ends, respectively), and left there for 20 days. At the end of this period the furnace was switched off and the crystals extracted from the tube. The basic characterization of the structural and magnetic properties of MnPS$_3$ crystals grown in this way is described below in the supplementary notes 2.1 and 2.2.

### 1.2. Device Fabrication

The assembly of the MnPS$_3$ tunnel barrier devices start with the preparation of suitable exfoliated crystals of few-layer graphene (FLG, typically 5-to-10 layer thick), hexagonal boron nitride (hBN) and MnPS$_3$ on silicon substrates covered with 300 nm SiO$_2$. The exfoliation procedure –as well as the assembly of the tunnel barriers – was performed in a glove box filled with nitrogen gas (with oxygen and water concentration in the sub-ppm level), to avoid degradation of the MnPS$_3$ layers. A by-now common temperature-controlled pick-up and release technique (based on a stack of PC polymer on a PDMS support) was employed to assemble the FLG/MnPS$_3$/FLG structures, and to encapsulate them between hBN crystals. The assembly of the heterostructure relies on three steps: the pick-up, the junction formation and the release (Fig S1 (a-c)). The pick-up process consists in putting the PC/PDMS stack in contact with the flake, heating it up to approximatively 130°C and then cooling down back to 100°C or lower temperature. Steadily removing the PC/PDMS stack allow then to lift the flake from substrate, leading to a PC/PDMS/Flake stack on one side and the substrate on the other side. The junction formation phase is composed of several pick-up steps, stacking the various flakes in the desired orientation using a same PC/PDMS support. The release phase is the last step of the assembly where the complete structure is then transferred on the final substrate. The PC/PDMS/Structure stack is again put in contact with the substrate, but in this case the temperature is increased to 180°C to separate the PC film from the PDMS leaving the whole structure onto the final substrate covered by the PC film which is then removed by immerging the sample into chloroform. Note that both the top and the bottom hBN flakes extend over the entire MnPS$_3$ layer to protect it from degradation, so that the structures are stable and can be taken out of the glove box. At this stage electron-beam lithography, electron beam evaporation, etching of hBN (to expose the FLG contacts away from MnPS$_3$ layer) and lift-off were then used to define metal electrode contacts.

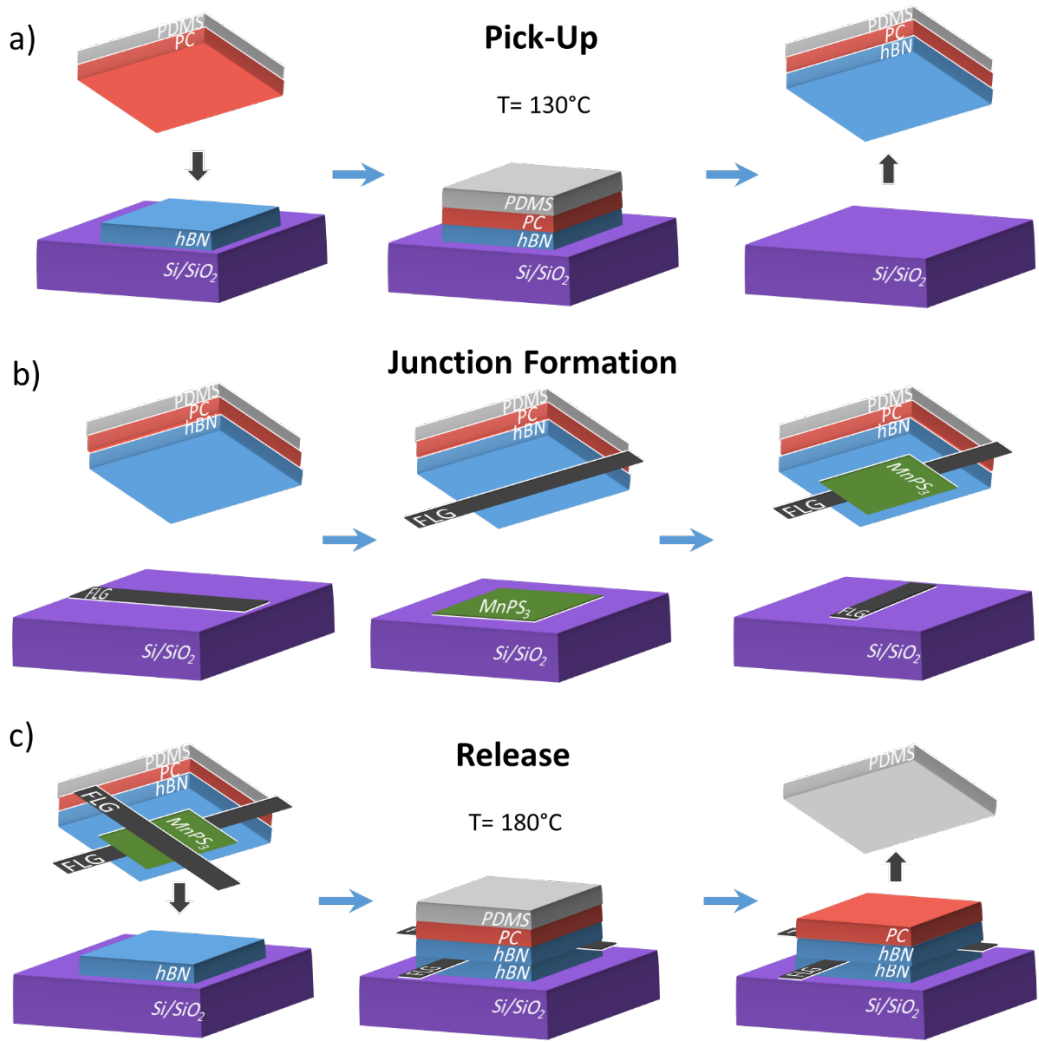

**Fig. S1. Device fabrication process.** Schematic of the layer-by-layer assembly of MnPS$_3$ tunnel barrier devices using PDMS/PC stack. The procedure is composed of three phases: the pick-up (a), the junction formation (b) and the release (c). In the pick-up and the release phases, the whole stack is heated at respectively 130°C and 180°C. The structure is built by means of several pick-up and a final release, where each flake is carefully aligned to ensure that the transport occurs through the MnPS$_3$ crystal and that the hBN flakes protect it from degradation.

### 1.3. Electrical Measurement

Transport measurements were performed in different cryostats, all equipped with superconducting magnets. Depending on the cryostat, the lowest temperature that could be reached was either 0.25 K or 1.5 K, and the highest magnetic field that could be applied was either 12 T or 14 T. To measure the *I-V* characteristics of the tunnel barrier and their magnetoresistance, the bias voltage was applied using either a Keithley 2400 source unit or a home-made low noise voltage source. The current and voltage signals were amplified with home-made low noise amplifier, and the amplified signals were recorded with an Agilent 34410A digital multimeter unit.

## 2. Supplementary Notes

### 2.1. Structural and stoichiometric characterization of $MnPS_3$ crystals

The quality of the $MnPS_3$ crystals was investigated by X-ray diffraction (XRD) and by energy-dispersive X-ray (EDX) spectroscopy. The XRD results showing only one set of narrow [001] reflections (Fig. S2) prove a good crystalline quality of the material. EDX analysis were performed in a scanning electron microscope (SEM). The SEM images show crystals with clean surfaces and sharp angles, and the EDX analysis confirms that the crystals have a homogeneous composition with the expected 1:1:3 elemental ratio of Mn, P and S atoms.

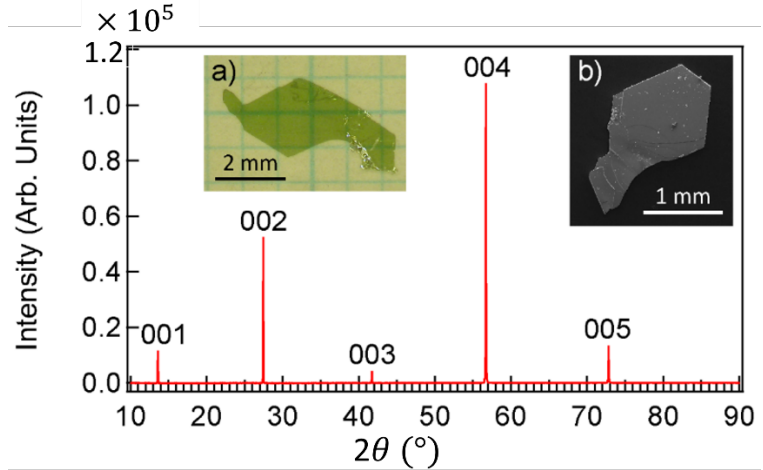

**Fig. S2. Characterizations of bulk MnPS$_3$.** X-ray diffraction patterns from the cleavage plane of a MnPS$_3$ crystal. Insets (a) and (b) show optical microscope (left inset) and scanning electron microscope (right inset) images of two different MnPS$_3$ crystals.

### 2.2. Temperature dependent magnetic susceptibility of bulk MnPS$_3$

The magnetic response of the MnPS$_3$ crystals grown in our laboratory was characterized by magnetization and magnetic susceptibility measurements (see also Fig. 1(b) in the main text) and compared to previous studies on bulk MnPS$_3$[56-58]. The temperature dependence of the magnetic susceptibility measured with an in-plane and out-of-plane field (see, respectively, the blue and red curves in Fig. S3) is identical in the two cases as $T$ is lowered from room temperature down to 78 K. The in-plane and out of plane susceptibilities start to differ for $T$ below the Néel temperature of bulk MnPS$_3$ crystals, $T_N$ = 78 K (indicated by the green arrow in Fig. S2). The observed behavior is fully in line with earlier studies reported in the literature (referred to in the main text).

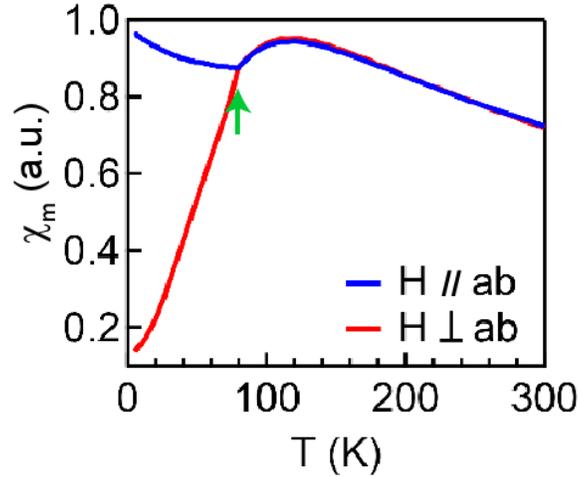

**Fig. S3. Identification of the transition temperature to the antiferromagnetic states in bulk MnPS$_3$.** Temperature dependence of the magnetic susceptibility of bulk MnPS$_3$ crystals with the magnetic field applied in-plane (blue curve) and perpendicular to the plane (red curve). The green arrow points to the temperature below which the two curves start to deviate one from the other, which allows us to identify the Néel temperature of bulk MnPS$_3$, $T_N$ = 78 K.

### 2.3. Thickness measurement of thin MnPS$_3$ crystals

For MnPS$_3$ crystals, systematic analysis of optical contrast does not give reliable estimation of the layer number. Thus, we used atomic force microscopy to measure the thickness of MnPS$_3$ flakes after the encapsulation (Fig S4 (a)), from which we identify the number of layers by extracting the height profile across the crystal step edge (FigS4 (b)).

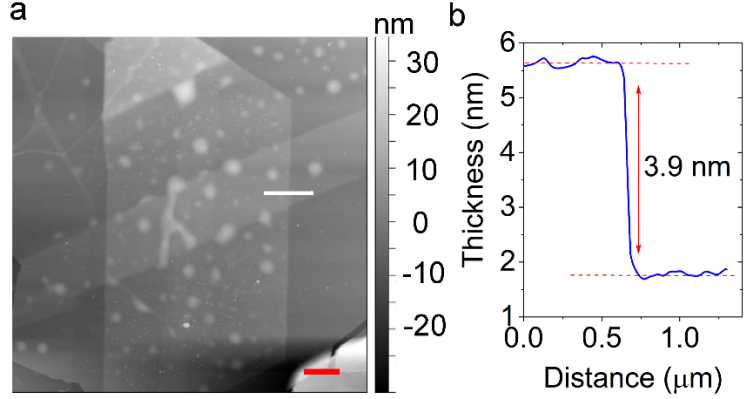

**Fig. S4. Atomic force microscopy measurement of a 6L MnPS$_3$ tunneling device.** (a) Topographic image of an encapsulated MnPS$_3$ vertical device fabricated as described in the main text (red scale bar is 1 µm). (b) Height profile across the edge of the barrier extracted along the white bar in (a) showing a step height of about 3.9 nm, which correspond to a 6L MnPS$_3$ crystal.

### 2.4. High-temperature magnetoresistance in monolayer and bilayer devices

As we discussed in the main text, thick MnPS$_3$ multilayers exhibit no magnetoresistance for $T > 80$ K, i.e., when the temperature is larger than the Néel bulk temperature, whereas in monolayer and bilayer some magnetoresistance is present even at these higher temperatures. We find, however, that for $T > 80$ K and up to the highest temperature of our measurements (120 K), the magnetoresistance observed in mono and bilayer MnPS$_3$ does not depend on $T$. The occurrence of the magnetic transition in mono- and bilayer can therefore be detected by looking at the temperature below which the magnetoresistance starts to be temperature dependent. In practice, as we discussed in the main text, to eliminate the effect of this $T$-independent background magnetoresistance, we subtract from the magnetoresistance measured at any given temperature, the same quantity measured at $T = 120$ K.

In Fig. S4 (a) and (b) we plot the magnetoresistance of mono- and bilayer MnPS$_3$ for $T$ between 80 and 120 K, to show that it is indeed temperature independent. This background magnetoresistance is likely coming from the few-layer graphene electrodes that give a measurable signal only in mono- and bilayers because the resistance of these thin tunnel barrier is much smaller than that of the thick MnPS$_3$ multilayers. Indeed, consistently with this statement, the magnetoresistance background of bilayer MnPS$_3$ is smaller than that of monolayer MnPS$_3$ (see Fig. S5(a) and (b)).

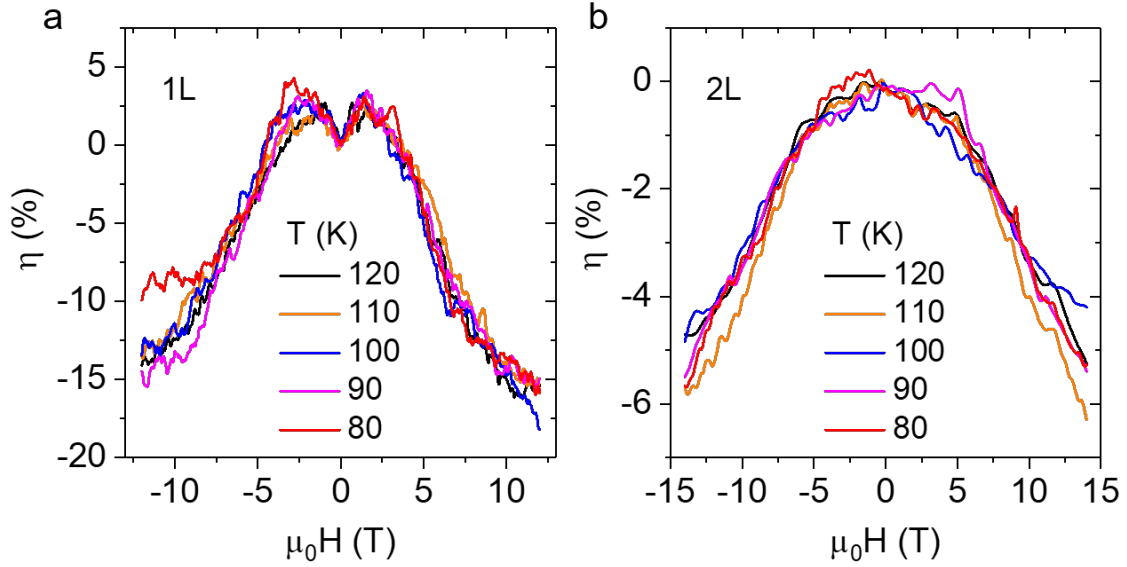

**Fig. S5. Magnetoresistance background in mono and bilayer MnPS$_3$ devices.** Magnetoresistance η of a (a) monolayer and a (b) bilayer tunnel junction measured at different temperatures above $T_N$. Within the noise, the signal does not depend on temperature for $T$ between 80 and 120 K.

### 2.5. First-principles calculations

First-principles calculations have been performed using the Quantum ESPRESSO suite of codes[59,60] within the generalized-gradient approximation of density functional theory as parameterized by Perdew, Burke, and Ernzerhof[61]. Structural relaxations have been performed assuming an antiferromagnetic configuration of spin on Mn atoms by minimizing all atomic forces and unit cell stress using the Broyden–Fletcher–Goldfarb–Shanno algorithm, until forces and stresses fall below $10^{-3}$ eV/Å and 0.5 kbar, respectively. The Brillouin zone has been sampled using a Γ-centered 8x8x1 Monkhorst-Pack grid, while electron-ion interactions have been approximated using pseudopotentials from the Standard Solid State Pseudopotential library[62] (v.1.0) with a cutoff energy of 70 Ry for wavefunctions and 840 Ry for the charge density. In structural relaxations spin-orbit coupling has been neglected and introduced only when computing band structures by using fully-relativistic optimized norm-conserving Vanderbilt (ONCV) pseudopotentials[63] from the Pseudo-Dojo library[64], with a cutoff of 80 Ry on wavefuntions. Spurious effects of the presence of artificial periodic replicas of the two-dimensional layer (inherent to the use of a plane-wave basis) have been avoided by adopting a real-space cutoff on Coulomb interactions in the direction orthogonal to the layers[65].

**Supplementary references:**

56      Okuda, K. *et al.* Magnetic Properties of Layered Compound MnPS$_3$. *Journal of the Physical Society of Japan* **55**, 4456-4463, (1986).
57      Joy, P. & Vasudevan, S. Magnetism in the layered transition-metal thiophosphates MPS$_3$ (M= Mn, Fe, and Ni). *Physical Review B* **46**, 5425, (1992).
58      Long, G. *et al.* Isolation and Characterization of Few-Layer Manganese Thiophosphite. *ACS Nano* **11**, 11330-11336, (2017).
59      Giannozzi, P. *et al.* QUANTUM ESPRESSO: a modular and open-source software project for quantum simulations of materials. *Journal of Physics: Condensed Matter* **21**, 395502, (2009).
60      Giannozzi, P. *et al.* Advanced capabilities for materials modelling with Quantum ESPRESSO. *Journal of Physics: Condensed Matter* **29**, 465901, (2017).
61      Perdew, J. P., Burke, K. & Ernzerhof, M. Generalized Gradient Approximation Made Simple. *Physical Review Letters* **77**, 3865-3868, (1996).
62      Prandini, G., Marrazzo, A., Castelli, I. E., Mounet, N. & Marzari, N. Precision and efficiency in solid-state pseudopotential calculations. *npj Computational Materials* **4**, 72, (2018).
63      Hamann, D. R. Optimized norm-conserving Vanderbilt pseudopotentials. *Physical Review B* **88**, 085117, (2013).
64      van Setten, M. J. *et al.* The PseudoDojo: Training and grading a 85 element optimized norm-conserving pseudopotential table. *Computer Physics Communications* **226**, 39-54, (2018).
65      Sohier, T., Calandra, M. & Mauri, F. Density functional perturbation theory for gated two-dimensional heterostructures: Theoretical developments and application to flexural phonons in graphene. *Physical Review B* **96**, 075448, (2017).